\documentclass[%
 reprint,
superscriptaddress,
 nobibnotes,
 amsmath,amssymb,
 aps, prl, reprint
]{revtex4-2}

\usepackage{graphicx}
\usepackage{dcolumn}
\usepackage{bm}

\usepackage{xcolor}
\usepackage{lineno}
\usepackage{float}
\usepackage{afterpage}

\usepackage{parskip}
\usepackage{lipsum}

\usepackage{amssymb,amsmath,amstext,amsthm}
\usepackage[utf8]{inputenc}
\usepackage[linesnumbered,ruled]{algorithm2e}
\usepackage{url}
\usepackage{soul}
\usepackage[normalem]{ulem}
\usepackage{multirow, makecell}
\usepackage{booktabs}
\usepackage{subfigure}
\usepackage{array}
\usepackage[colorinlistoftodos]{todonotes}
\usepackage{orcidlink}

\definecolor{ablightblue}{HTML}{99b6ff}

\makeatletter
\def\@fnsymbol#1{\ensuremath{\ifcase#1\or
   *\or \dagger\or \ddagger\or \mathsection\or \mathparagraph\or \|\or
   **\or \dagger\dagger\or \ddagger\ddagger\or
   \mathsection\mathsection\or \mathparagraph\mathparagraph\or \|\|\or
   ***\or \dagger\dagger\dagger\or \ddagger\ddagger\ddagger\or
   \mathsection\mathsection\mathsection\or \mathparagraph\mathparagraph\mathparagraph\or \|\|\|\or
   ****\or \dagger\dagger\dagger\dagger\or \ddagger\ddagger\ddagger\ddagger
   \else\@ctrerr\fi}}
\makeatother

\usepackage{hyperref}
\hypersetup{
    hidelinks,               
    colorlinks=true,         
    citecolor=teal,          
    linkcolor=teal,           
    urlcolor=teal           
}

\newcommand{\orcidauthor}[2]{%
    \author{#1%
    \if\relax\detokenize{#2}\relax
    \else\,\orcidlink{#2}%
    \fi}%
}

\begin{document}

\title{A Novel Approach to Short Baseline Oscillation Searches Using Neutrino Tagging with nuSCOPE}

\orcidauthor{Adrien Blanchet}{0000-0002-4992-0161}
\email{adrien.blanchet@cern.ch}
\affiliation{Laboratoire de Physique Nucléaire et de Hautes Énergies (LPNHE), CNRS/IN2P3, Sorbonne Université, Université Paris Cité, 4 place Jussieu, 75005 Paris, France}
\affiliation{European Organization for Nuclear Research (CERN), EP-NU, 1211 Geneva 23, Switzerland}

\orcidauthor{César Jesús-Valls}{0000-0002-0154-2456}
\email{cesar.jesus-valls@cern.ch}
\affiliation{European Organization for Nuclear Research (CERN), EP-NU, 1211 Geneva 23, Switzerland}

\orcidauthor{Animesh Chatterjee}{0000-0002-2935-0958}
\affiliation{European Organization for Nuclear Research (CERN), EP-NU, 1211 Geneva 23, Switzerland}

\orcidauthor{Stephen Dolan}{0000-0002-2410-6550}
\affiliation{European Organization for Nuclear Research (CERN), EP-NU, 1211 Geneva 23, Switzerland}

\orcidauthor{Pierre Granger}{0000-0002-8333-4393}
\affiliation{European Organization for Nuclear Research (CERN), EP-NU, 1211 Geneva 23, Switzerland}

\orcidauthor{Laura Munteanu}{0000-0002-2074-8898}
\affiliation{European Organization for Nuclear Research (CERN), EP-NU, 1211 Geneva 23, Switzerland}

\begin{abstract}
    We present the first study of short-baseline neutrino oscillation searches using a tagged neutrino beamline,
    taking the proposed nuSCOPE facility at CERN as a benchmark.
    In this Letter we demonstrate that tagged neutrino beams, where the neutrino flavor, energy,
    and propagation distance are determined with exceptional event-by-event precision,
    provide a new experimental approach to search for non-standard neutrino oscillations.
    We evaluate the sensitivity to sterile-neutrino-induced oscillations in the $\nu_\mu$ disappearance,
    $\nu_\mu \rightarrow \nu_e$ appearance, and $\nu_e$ disappearance channels,
    demonstrating the ability to probe multiple flavor transitions
    and both appearance and disappearance modes within a single experiment.
    Our results show that tagged beams enable sensitivity improvements to mass-squared splittings spanning several orders
    of magnitude while substantially reducing the dependence on neutrino flux predictions that limits
    conventional searches.
    We find that nuSCOPE can probe a broad region of parameter space motivated by existing anomalies
    and extend coverage into previously unexplored territory,
    demonstrating the strong potential of tagged neutrino beams for precision oscillation physics.
\end{abstract}

\maketitle

{\em Introduction.}
While the three-flavor PMNS framework has achieved remarkable success in describing
neutrino oscillations across a wide range of sources, detection technologies,
and experimental configurations~\cite{ParticleDataGroup:2024cfk}, several
experimental anomalies remain~\cite{Acero:2022wqg}.
Two broad interpretations of these observations have been put forward.

The first attributes them to the complexity of modeling
neutrino production and interaction mechanisms.
In experiments using nuclear reactors, neutrino flux predictions must
account for uncertainties in the fission yields and beta-spectrum conversion
methods~\cite{Perisse:2023efm}, while measurements of the neutrino energy depend
directly on nuclear-structure uncertainties during calibration~\cite{Elliott:2023xkb}.
In atmospheric and accelerator neutrino experiments, the flux prediction depends
critically on the modeling of charged hadron production~\cite{NA61SHINE:2018rhe,
Honda:2011nf}, and the neutrino energy reconstruction is highly sensitive to
poorly understood nuclear effects in neutrino-nucleus
interactions~\cite{Dolan:2026nlr}. Limitations in
the modeling of these processes may affect the predicted neutrino spectra and rates.

The second associates the anomalies with the possible existence of at least one
additional neutrino mass state, most commonly a sterile neutrino that does not
participate in weak interactions but mixes with the three known active flavors.
Such a state would introduce new oscillation frequencies with large mass-squared
differences ($\Delta m^2 \sim 1 - 10^4$ eV$^2$), leading to short-baseline
oscillations observable as distortions in the energy spectra or as
appearance/disappearance signals.

In this Letter, we demonstrate for the first time the potential of a tagged
neutrino beamline, namely one instrumented to provide exceptional event-by-event
knowledge of the initial neutrino flavor, energy ($E_\nu$), and propagation length
($L_\nu$), to probe oscillations at short baseline. We present the expected
sensitivities of the recently proposed nuSCOPE experiment at CERN~\cite{Acerbi:2025wzo}
and show how this approach can probe a vast region of the sterile oscillation
parameter space, covering existing anomalies while extending to previously
uncharted eV-scale territory. This concept offers two further advantages: (1) the
simultaneous study of disappearance and appearance channels for both the $\nu_e$
and $\nu_\mu$ flavors, and (2) the ability to compare measurements and test the
consistency of oscillations between neutrinos and antineutrinos. Unlike existing
experiments, such a multichannel search enables rigorous cross-checks of the
results and a detailed exploration of sterile neutrino phenomenology should
new-physics signatures emerge.

\bigbreak
{\em Experiment Description.}
The proposed nuSCOPE facility at CERN~\cite{Acerbi:2025wzo} would use a horn-less
beamline driven by 400 GeV protons extracted from the Super Proton Synchrotron
(SPS) to produce a narrow-band beam of $\pi^+$ and $K^+$ mesons at a central
momentum of 8.5 GeV/$c$, focused by a static system of quadrupoles and dipoles. To
improve control of neutrino production and interactions, nuSCOPE would combine two
complementary techniques. The first, developed by the ENUBET
collaboration~\cite{ENUBET:2023hgu}, consists of instrumenting the walls of the
40-meter decay tunnel with a modular sampling calorimeter and photon-veto system,
operated in a slow-extraction mode (4.8–9.6 s spills) to minimize particle pile-up.
This would allow direct monitoring of the charged leptons ($e^+$ and $\mu^+$)
produced alongside neutrinos in meson decays, constraining the $\nu_e$ and $\nu_\mu$
fluxes to 1\% precision, an order of magnitude better than in conventional beams.
The second technique, pioneered by the NuTag
collaboration~\cite{Baratto-Roldan:2024bxk}, employs silicon pixel tracking
detectors with sub-100 ps time resolution, positioned before and after the decay
tunnel. These spectrometers measure the momentum and trajectory of both the parent
mesons and the daughter muons from $\pi^+ \rightarrow \mu^+ \nu_\mu$ and $K^+
\rightarrow \mu^+ \nu_\mu$ decays. When combined with a detector located 25 meters
downstream, this setup would enable event-by-event tagging of neutrino
interactions: each observed neutrino could be uniquely associated with its parent
decay, and its energy determined from two-body kinematics with sub-percent
precision. The neutrino-tagging concept was recently demonstrated by the NA62
experiment~\cite{NA62:2024xzj}.

Lastly, nuSCOPE would include a dedicated neutrino detector downstream of the beam
dump. The reference detector is a LAr-TPC with a 500-ton fiducial mass and 300 ps
timing resolution, building on technology developed for the ProtoDUNE detectors at
CERN's Neutrino Platform~\cite{DUNE:2021hwx}. Its $4 \times 4$ m$^2$ front face and
22.3 m length would provide significant acceptance for the neutrino beam while
maintaining excellent spatial and energy resolution. The facility is designed to
collect $10^4$ $\nu_e$ and $10^6$ $\nu_\mu$ charged-current events with $1.4 \times
10^{19}$ protons on target over approximately five years of operation.

\bigbreak
{\em Methodology.}
We assess the sterile neutrino sensitivity of nuSCOPE utilising neutrino events generated with the
nuSCOPE simulation pipeline described in Ref.\cite{Acerbi:2025wzo}.
The analysis considers both disappearance ($\nu_\mu \rightarrow \nu_\mu$ and $\nu_e \rightarrow \nu_e$) and appearance ($\nu_\mu \rightarrow \nu_e$)
oscillation channels over baselines ranging from the beam dump entrance (25~m) to the downstream end of the detector
(87~m), corresponding to the 40~m decay tunnel plus the 22~m detector length.
As these baselines are much shorter than long baseline configurations, the oscillation probabilities
can be simplified by using the two flavours approximation.
For $\nu_\alpha \rightarrow \nu_s$ disappearance, the survival probability is $P_{\alpha\alpha} = 1 - \sin^2(2\theta_{\alpha s}) \sin^2(1.267 \Delta m^2 L/E)$,
where $\Delta m^2$ is in eV$^2$, $L$ in km, and $E$ in GeV.
For $\nu_\mu \rightarrow \nu_e$ appearance, $P_{\mu e} = \sin^2(2\theta_{\mu e}) \sin^2(1.267 \Delta m^2 L/E)$.

The Monte Carlo sample is built from the nuSCOPE flux simulation described in Ref.~\cite{Acerbi:2025wzo},
which uses \textsc{BDSim}~\cite{Nevay:2018zhp} for the beam transport and \textsc{Geant4}~\cite{GEANT4:2002zbu} for the beamline
instrumentation and detector geometry.
Neutrino interactions in the liquid-argon fiducial volume are generated with
\textsc{GENIE} using the AR23\_20i\_00\_000 configuration,
following Ref.~\cite{GENIE:2021npt}.
The predicted tagged $\nu_\mu$ event yield is obtained by applying
the tagging efficiencies from Ref.~\cite{Acerbi:2025wzo} to the
generated $\nu_\mu$ charged-current sample.
For five years of operation, we consider 760k tagged $\nu_\mu$
charged-current interactions and 12k untagged $\nu_e$ charged-current
interactions in the neutrino detector fiducial volume.
Here, ``tagged'' denotes the event-by-event association of a neutrino
interaction with its parent meson decay; no such association is used
for the $\nu_e$ disappearance sample.
No additional neutrino-detector reconstruction or selection efficiency is applied
in this study.
The projections thus assume unit detector efficiency; a geometry-dependent
reconstruction and selection assessment is left to future work.

\begin{figure}[!h]
    \centering
    {\includegraphics[width=0.99\linewidth]{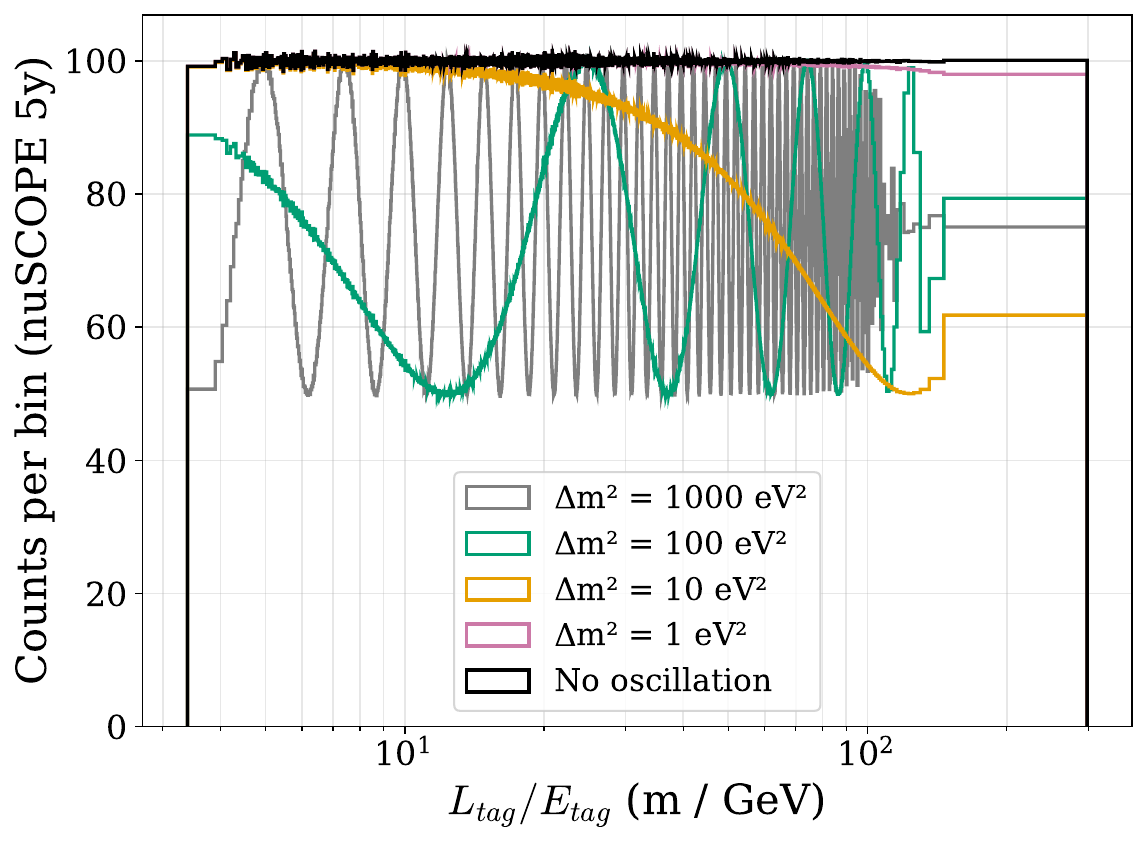}}
    \caption{
        Expected $\nu_\mu$ spectrum histograms with arbitrary $\sin^{2}2\theta_{24} = 0.5$
        for different values of $\Delta m_{42}^2$.
        For the lowest values of $\Delta m^2$ the oscillation pattern start to develop
        for the highest bins in $L_\text{tag}/E_\text{tag}$.
        On the other hand, for the high values of $\Delta m^2$ multiple oscillation maxima
        can be observed, thanks to the ability of nuSCOPE to provide an accurate
        event-by-event measurement of both the propagation distance and energy.
        The wide $L_\text{tag}/E_\text{tag}$ spectrum allows to statistically infer
        on the hypothesis of an oscillation towards a sterile mass state by only
        testing different shape distortions.
        For $\Delta m^2 > 1000\text{eV}^2$ the last bins in $L_\text{tag}/E_\text{tag}$
        average the oscillation patterns to $1/2$ of the oscillation amplitude.
        At very high oscillation frequencies $\Delta m^2 > 1\text{keV}^2$, every bin holds
        multiple oscillation maxima.
        The expected sensitivity then mainly relies on the accuracy knowledge of the
        absolute normalization.
    }
    \label{fig:numuSurvivalSpectra}
\end{figure}

As muon neutrinos are mostly produced via two body decays,
i.e. $\pi^+ \rightarrow \mu^+ + \nu_\mu$ and $K^+ \rightarrow \mu^+ + \nu_\mu$,
the tagging offers a direct probe of the neutrino energy.
The energy resolution has been studied in Ref.\cite{Acerbi:2025wzo}.
We reproduced the energy resolution as a function of the neutrino energy by $\sigma(E_\nu)/E_\nu = a/\sqrt{E_\nu} + b/E_\nu + c$,
with parameters $(a, b, c) = (0.347, 0.197, 0.008)$.
We sampled the reconstructed energy $E_\textrm{tag}$ of each event by throwing the values in a gaussian
centered around the true energy $E_\nu$ with a width of $\sigma(E_\nu)$.
This procedure introduced a natural smearing on the oscillation patterns in $L_\text{tag}/E_\text{tag}$.
We use a quantile-based adaptive binning in $L_\text{tag}/E_\text{tag}$, with an expected number of events
per bin corresponding to a 1\% statistical uncertainty.
This choice allows us to take advantage of the shape information with thinner bins that contain many events.
Fig.~\ref{fig:numuSurvivalSpectra} shows the expected Asimov spectrum histograms with different
choices of oscillation frequencies $\Delta m^2$.

We construct a $\chi^2$ test comparing the expected event distributions for various sterile neutrino oscillation predictions with respect to the no oscillation hypothesis. We perform the statistical tests by scanning a two-dimensional grid in $(\sin^2 2\theta, \Delta m^2)$ parameter space with
$\Delta m^2$ ranging from $10^{-2}$ to $10^{4}$~eV$^2$ (100 logarithmic steps)
and $\sin^2 2\theta$ from $10^{-3}$ to $1$ (100 logarithmic steps).
At each grid point, we compute $\Delta \chi^2$ relative to the null hypothesis and report sensitivity contours at 90\% CL and $2\sigma$ significance for two degrees of freedom.

For $\nu_\mu$ disappearance, we use a shape-only likelihood fit.
At each point in $(\sin^2 2\theta,\Delta m^2)$ space, the predicted bin contents $\mu_i$ are multiplied by a free overall normalization factor $\phi$,
which is profiled to match the total observed event rate.
The sensitivity therefore comes only from the relative distortions of the $L_{\mathrm{tag}}/E_{\mathrm{tag}}$ spectrum across bins, rather than from the absolute normalization.
This conservative choice isolates the sensitivity arising from spectral distortions alone,
which dominates in the eV$^2$ to keV$^2$ region considered here,
and avoids relying on absolute rate information for this conceptual demonstration.
At large $\Delta m^2$, the oscillation wavelength in $L/E$ becomes smaller than the effective binning scale, so the fast modulation is no longer physically resolved and the bin expectation tends toward the averaged limit $\langle \sin^2 \varphi \rangle = 1/2$.
With finite Monte Carlo statistics, however, bins spanning several oscillation periods can still exhibit residual fluctuations around this average. 
To suppress these numerical aliasing artifacts, we estimate for each bin $b$ the phase excursion
$\Delta \varphi_b = 1.267\,\Delta m^2\,\Delta (L/E)_b$ and progressively damp the oscillatory term toward $1/2$ with an exponential weight
$w_b = 1 - e^{-\Delta \varphi_b/(2\pi)}$, so that the transition to the averaged regime is controlled by the phase scale $2\pi$.
For $\nu_\mu \rightarrow \nu_e$ appearance, we use a Poisson likelihood
and show an idealized, statistics-only, zero-background projection.
Non-negligible backgrounds are nevertheless expected, notably from gammas
in neutral-current (NC) $\pi^0$ interactions, which tagging alone cannot reject.
Measurements on argon suggest an NC $\pi^0$-to-CC rate ratio of order
5--10\%, depending on the spectrum and final-state
definition~\cite{MicroBooNE:2022NCpi0,MicroBooNE:2019CCinclusive,ArgoNeuT:2017NCpi0}.
Assuming comparable tagging acceptance, this gives roughly
$(4$--$8)\times10^4$ NC $\pi^0$ events in our sample before electron-like selection.
A simulation study of ICARUS-type LAr-TPCs estimates an NC $\pi^0$
misidentification probability of $10^{-3}$ at approximately 90\% $\nu_e$
selection efficiency within the fiducial volume~\cite{Baibussinov:2009CERNPS}.
Used as an illustrative benchmark for nuSCOPE, this rejection would leave
40--80 background events, pending a detector-specific assessment.
At unit signal efficiency, including a fixed, known background of 75 events in the counting likelihood
weakens the limit on $\sin^2(2\theta_{\mu e})$ by somewhat less than one order
of magnitude relative to the displayed zero-background contour.

\begin{figure}
    \centering
    {\includegraphics[width=0.99\linewidth]{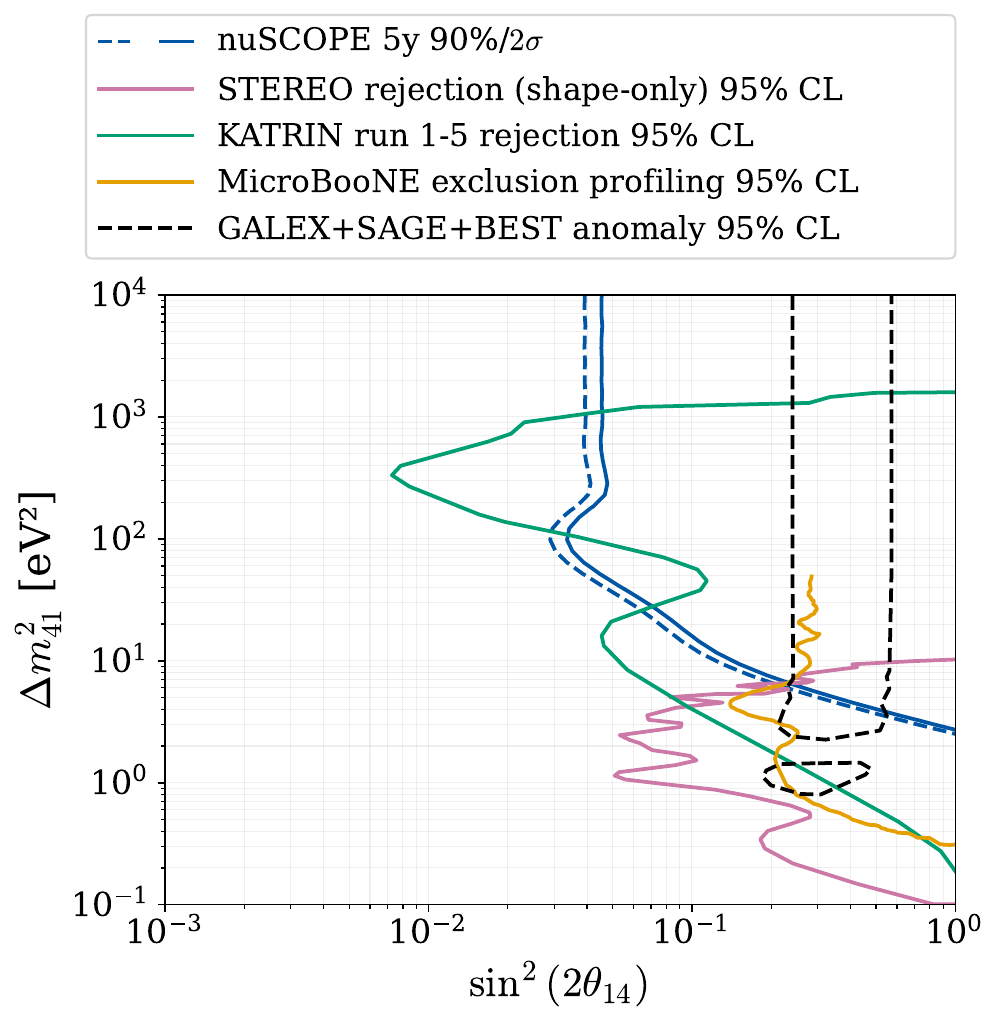}}
    \caption{
    Sensitivity to the disappearance channel $\nu_e \rightarrow \nu_e$.
    The nuSCOPE projection includes statistical uncertainties only,
    with the predicted normalization held fixed.
    The GALLEX+SAGE+BEST signal is the so called Galium anomaly or
    radioactive source anomalies from~\cite{Barinov:2021asz}.
    The \textsc{Stéréo} contour was extracted from~\cite{STEREO:2022nzk}.
    Electron antineutrinos were produced within a fission reactor with a
    typical energy of $E_\nu \sim 2-8$ MeV.
    Their exclusion contours were realized with the same methodology as our
    $\nu_\mu$ disappearance as they measured the antineutrino spectra for six
    different propagation lengthes.
    The KATRIN exclusion contour was taken from~\cite{KATRIN:2025lph}.
    The expected signal is a kink-like distortion in the tritium beta-decay
    endpoint spectrum.
    The MicroBooNE rejection contour is from~\cite{MicroBooNE:2025nll}.
    The oscillation signal would have been a rate deficit of the intrinsic $\nu_e$ component
    of the beam.
    }
    \label{fig:nueDisppearance}
\end{figure}

The $\nu_e$ disappearance analysis uses the untagged $\nu_e$ sample.
Since these neutrinos are predominantly produced in three-body kaon
decays, the two-body kinematic reconstruction used for the tagged
$\nu_\mu$ sample is not applicable.
We use the reconstructed outgoing-electron energy,
$E_{\mathrm{lep}}^{\mathrm{reco}}$, as the analysis observable.
The relationship between the true neutrino energy, $E_\nu^{\mathrm{true}}$,
and the true outgoing-electron energy, $E_{\mathrm{lep}}^{\mathrm{true}}$,
is provided by the \textsc{GENIE} interaction simulation, which accounts
for the energy transferred to the hadronic system.
To approximate the effects of electromagnetic-shower reconstruction,
we then sample $E_{\mathrm{lep}}^{\mathrm{reco}}$ from a Gaussian centered
on $E_{\mathrm{lep}}^{\mathrm{true}}$ with a width
$\sigma_{\mathrm{lep}} = 0.10 E_{\mathrm{lep}}^{\mathrm{true}}$.
This constant resolution is a simplified approximation to LAr-TPC
electromagnetic-shower reconstruction performance. For comparison,
the DUNE simulation reports an electron-energy resolution of approximately
$4\% \oplus 9\%/\sqrt{E_e/\mathrm{GeV}}$~\cite{DUNE:2020LongBaseline},
where $\oplus$ denotes addition in quadrature; our 10\% assumption is
therefore conservative for electron energies above approximately 1~GeV.
This benchmark resolution applies only to the electron energy;
$E_{\mathrm{lep}}^{\mathrm{reco}}$ is not a calorimetric reconstruction
of the full neutrino energy.
Oscillation probabilities are evaluated using $E_\nu^{\mathrm{true}}$,
and the resulting event weights are used to predict
the $E_{\mathrm{lep}}^{\mathrm{reco}}$ spectrum.
Since the absence of tagging prevents precise event-by-event $L$
determination, the sensitivity is largely driven by the total event rate.
Applying this smearing has a negligible effect on the sensitivity.
The projection in Fig.~\ref{fig:nueDisppearance} includes statistical
uncertainties only, with the flux and interaction-rate normalization
held fixed; no flux, cross-section, or detector systematic uncertainties
are included.
Measurements of kaon decays with the tagging spectrometers are expected
to constrain the parent kaon population and hence the $\nu_e$
flux~\cite{Acerbi:2025wzo}.
The precision achievable with this method
remains to be quantified; the present projection therefore represents
the statistical limit rather than a sensitivity including these uncertainties.

\bigbreak

\begin{figure}
    \centering
    {\includegraphics[width=0.99\linewidth]{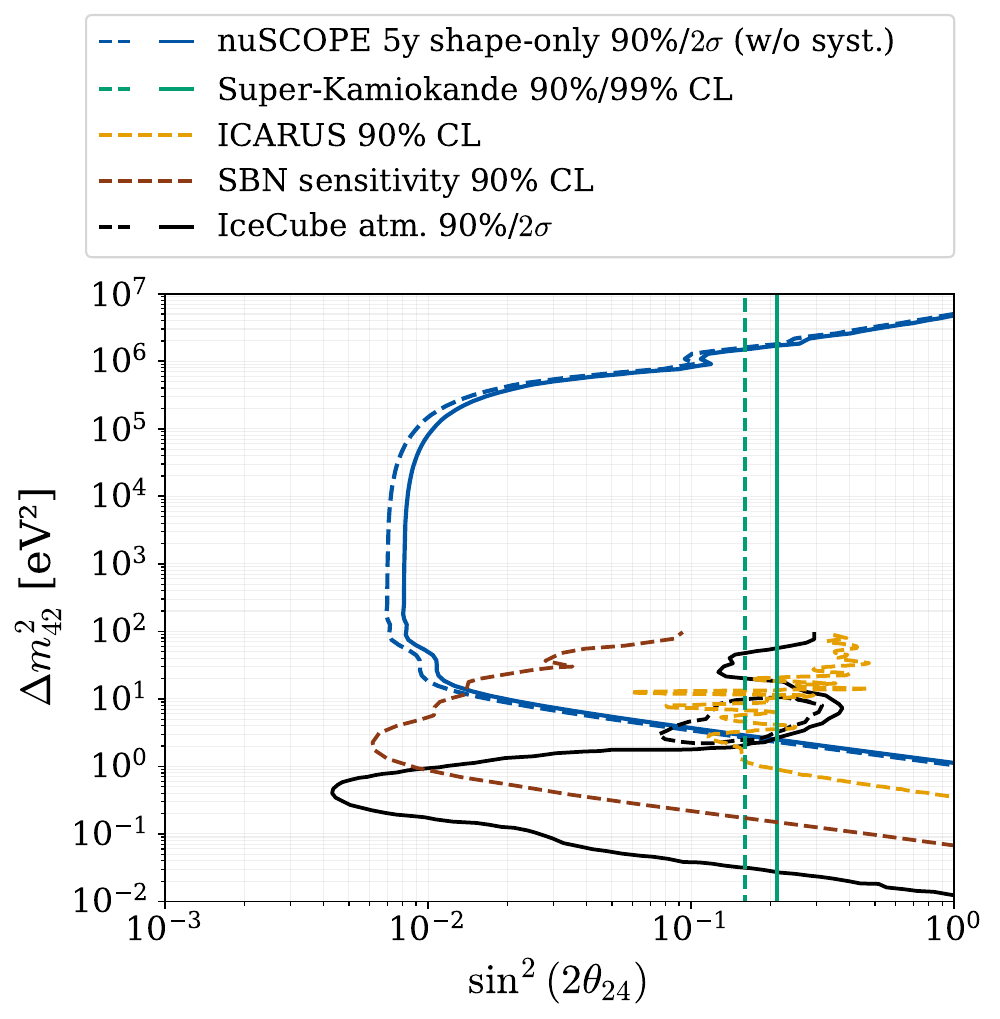}}
    \caption{
    Sensitivity to the disappearance channel $\nu_\mu \rightarrow \nu_\mu$.
    The projected SBN sensitivity at 90\% CL is from Ref.~\cite{SBND:2025sensitivity}.
    The right side of each open concours represents the exclusion area.
    The IceCube contours were taken from~\cite{IceCubeCollaboration:2024nle} where
    they used high-energy atmospheric neutrino ranging from $1$ to $100\text{ TeV}$.
    The Super-Kamiokande limits are extracted from~\cite{Super-Kamiokande:2014ndf},
    where GeV atmospheric neutrinos were used.
    }
    \label{fig:numuDisppearance}
\end{figure}

\begin{figure}
    \centering
    {\includegraphics[width=0.99\linewidth]{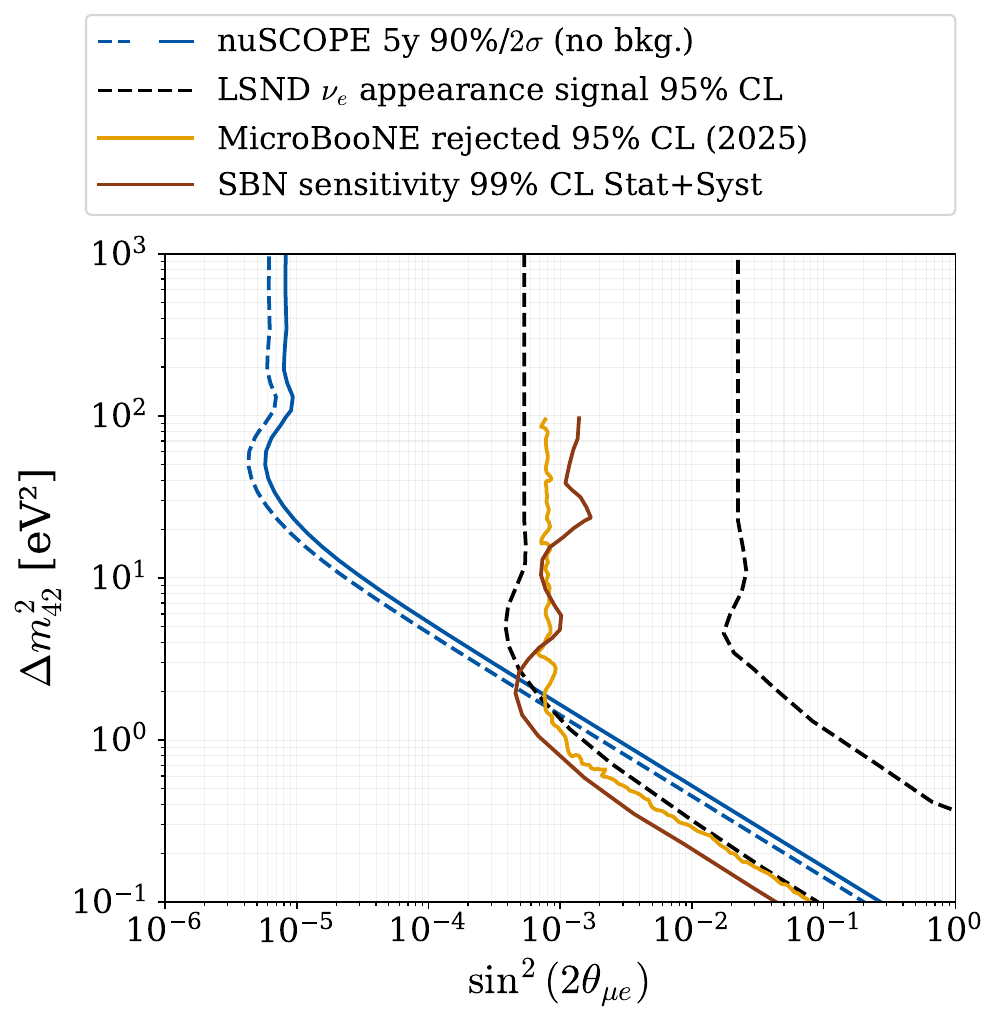}}
    \caption{
    Sensitivity to the appearance channel $\nu_\mu \rightarrow \nu_e$.
    The projected SBN sensitivity at 99\% CL, including statistical and systematic uncertainties, is from Ref.~\cite{SBND:2025sensitivity}.
    The nuSCOPE projection assumes unit detector efficiency, zero background,
    and statistical uncertainties only.
    The LSND positive signal contour was extracted from~\cite{LSND:2001aii} while
    the MicroBooNE rejection contour is from~\cite{MicroBooNE:2025nll}.
    LSND signal comes from muon decay at rest that produces $\bar{\nu}_\mu$ at
    $E_\nu \sim 20-60$ MeV. It tested the appearance of $\bar{\nu}_e$.
    MircoBooNE neutrinos are produced by the Booster Neutrino Beam by pion decays.
    The $\nu_\mu$ are emitted with a typical energy of $E_\nu \sim 0.5-2$ GeV,
    while the MicroBooNE detector is looking for interactions of $\nu_e$
    that oscillated.
    }
    \label{fig:nueAppearance}
\end{figure}

{\em Results.}
The projected nuSCOPE sensitivities for $\nu_\mu$ disappearance, $\nu_e$ appearance and $\nu_e$ disappearance are
presented in Figures~\ref{fig:numuDisppearance},~\ref{fig:nueAppearance}~and~\ref{fig:nueDisppearance} respectively.

Thanks to the event-by-event estimation of $L_{tag}/E_{tag}$, nuSCOPE would be able to probe
$\Delta m^2_{42}$ over
six orders of magnitude through spectral shape distorsions only.
$\Delta m^2_{42}$ sensivity is ranging from $1\text{ eV}^2$ to $1\text{ keV}^2$, for which previous
experimental constraints are limited by the overall normalization of the $\nu_\mu$ flux and cross-section
models.
In this analysis we purposely ignored the flux uncertainty constraints that would provide additional
sensitivity in the $\Delta m^2_{42} > 1\textrm{ keV}^2$ region.
The accuracy of nuSCOPE for constraining the flux and cross-section systematics are expected to
be the most precise thanks to the monitored muon flux in the decay tunnel.

Under these idealized assumptions, Fig.~\ref{fig:nueAppearance} shows sensitivity to $\sin^2(2\theta_{\mu e})$
in a phase-space overlapping with almost the totality of the LSND contours.
The tagging provides a very stringent constraint on the $\nu_e$ appearance as
an electron shower in the neutrino detector has to be associated with a matching meson
decay.

Finally, under the statistical-only assumptions used here,
Fig.~\ref{fig:nueDisppearance} shows that nuSCOPE would probe
the higher-$\Delta m^2_{41}$ portion of the gallium anomaly region.
The sensitivity reaches mixing amplitudes $\sin^2(2\theta_{14})$
of a few $10^{-2}$ for $\Delta m^2_{41} \gtrsim 100$~eV$^2$,
while the low-$\Delta m^2_{41}$ region around $1$~eV$^2$
remains beyond reach.

\bigbreak
{\em Conclusion.}
Tagged neutrino beamlines open a new experimental avenue for short-baseline oscillation searches.
By providing event-by-event information on the neutrino flavor, energy, and production point,
a facility such as nuSCOPE enables a precise reconstruction of the oscillation variable $L_\nu/E_\nu$.
This capability allows the exploration of oscillation patterns over a wide range of mass-squared
differences while significantly reducing the systematic uncertainties traditionally associated with
neutrino flux predictions and interaction modeling.

The sensitivity studies presented in this work demonstrate that nuSCOPE can probe a large fraction of the
parameter space motivated by current anomalies, while extending the search to previously unexplored
regions of the eV-scale sterile neutrino landscape.
The simultaneous access to multiple appearance and disappearance channels for both $\nu_e$ and $\nu_\mu$
flavors further provides powerful internal consistency tests that are difficult to achieve in
conventional beam experiments.

Beyond sterile neutrino searches, the nuSCOPE concept illustrates the broader physics potential of
tagged neutrino beams for precision oscillation measurements and neutrino interaction studies.
This approach represents a promising step toward a new generation of neutrino experiments with unprecedented control of the neutrino initial state.


\makeatletter
\let\pre@bibdata\@empty
\makeatother
\bibliographystyle{apsrev4-1}
\bibliography{biblio}

\end{document}